\documentclass[conf]{IEEEtran}
\usepackage{graphicx}

\begin{document}

\title{The Internet of Things: Perspectives on Security from RFID and WSN}

\author{A. Shah, A. Pal and H. B. Acharya\\
IIIT Delhi, India\\
acharya@iiitd.ac.in}

\maketitle

\begin{abstract}
  A massive current research effort focuses on combining pre-existing
  ``Intranets'' of Things into one Internet of Things. However, this
  unification is not a panacea; it will expose new attack surfaces and
  vectors, just as it enables new applications. We therefore urgently
  need a model of security in the Internet of Things. In this regard,
  we note that IoT descends directly from pre-existing research (in
  embedded Internet and pervasive intelligence), so there exist
  several bodies of related work: security in RFID, sensor networks,
  cyber-physical systems, and so on. In this paper, we survey the
  existing literature on RFID and WSN security, as a step to compiling
  all known attacks and defenses relevant to the Internet of Things.
\end{abstract}

\section{Introduction}

In the context of the current interest in the Internet of Things, it is instructive to remember previous attempts to build RFID-tagged supply chains, smart spaces (such as smart homes and offices, as well as smart cars), and automatic control systems for systems ranging from engines to power grids. Despite the extensive security literature developed in each of these areas, a general model of security remains elusive. This is a serious problem, as seen in recent breaches such as Slammer and Duqu; the Internet of Things is currently being developed without a clear understanding of the security of cyber-physical systems, leave alone an appreciation of the assumptions (of scale, name semantics, etc.) that we break when we connect them to the Internet.

Clearly, the extremely wide variety of use cases for the Internet of Things makes it impossible to have a single ``silver bullet'' solution to security. As a starting point, we believe it is essential to collect the known attack and defence models for the devices, services and networks that form IoT components or connectors. In this paper, we start by compiling a short survey of research in security for radio-frequency identification tags, and for wireless sensor networks. We summarize and classify existing attacks into three areas - privacy, authentication, and availability; survey the available solutions; and identify some shortcomings and open questions in the current state-of-the-art. 

\section{RFID security}

The earliest plans to define an ``Internet of Things'' involved the widespread use of RFID tags - small, usually passive electronic tags, remotely read by their modulation of radio-frequency waves.  These tags are both ubiquitous and cheap, and offer a direct replacement for pre-existing tag technology such as bar codes. The primary use is to automate the unique identification of objects, allowing them to be tracked, and to enable them to interact with their environment in ``smart'' ways; for example, a woolen garment might set a washing machine to the proper setting automatically.

RFID tags, however, may also be active and battery-powered. We propose, after Juels et al \cite{juels} that the primary distinguishing factor of a tag is that it serves to identify itself (and the object or person it is tagging), and thus excludes mobile phones and sensor networks. (Some ``RFID tags'', which can report if a box is opened or closed, for example, blur the lines.)  ISO 14443 and 15693 define proximity (10 cm) and vicinity (1 m) RFID devices for High-Frequency band tags; ISO 18000 specifies protocols for multiple bands (LF, HF, UHF). ISO 18092 introduces the concept of NFC - devices that can operate as both readers and tags, as required.

\subsection{Attacks on RFID}

RFID infrastructure is subject to three principal attacks: attacks on \emph{privacy}, on \emph{authentication}, and on \emph{availability}. We now explore these attacks, further classifying them by whether they target the tags and readers (edge hardware), the communication system, or the backend (database, middleware etc.) \cite{mit}. 

\subsubsection{Privacy}

The first essential question with regard to privacy is whose privacy, and what constitutes an attack on it. Traditionally, two forms of privacy are considered: the privacy of individual consumers (e.g. not being able to list a person's prostheses and implants by bringing a scanner near them), and the privacy of logistics (not being able to map out supply chains). However, we suggest that the Juels\cite{juels} classification of privacy threats into \emph{tracking} and \emph{inventorying} is more precise. 

As the name indicates, tracking involves developing a whole-life itinerary of a tag, by collecting readings at various points of its life cycle; this is particularly important considering there were proposals to put RFID tags in currency notes \cite{L41}. Collating data about the trajectories of multiple objects associated with a person provides much more information than, for example, cell phone GPS (which only tracks their current position). Tracking can combine several personas of one person, exposing dangerous or embarrassing information (e.g. exposing the personal life of an authority figure or celebrity); information about movements can be used to infer health or financial issues; relationships can be inferred (X and Y share clothing, they must be very close), and so on. 

Inventorying is a one-off read of a person or object to list their tags. Standardized ID formats, such as EPCglobal, disclose a great deal of information - manufacturer, product name, serial number, etc.; it has been speculated that, for example, RFID-enabled passports could be used to trigger smart bombs. (The profiling and selection of targets based on inventorying is called \emph{hotlisting}.) 

The clear distinction between the two attacks is that tracking can simply use the tag ID (or a \emph{constellation} - in absence of unique ID, a unique combination of IDs for tags travelling together) for monitoring, location disclosure, and so on. The semantic content of the tag is not needed. Hence, it is not enough for a tag to be encrypted or not linkable to ID; a stronger requirement is \emph{indistinguishability}, where the tag output cannot be distinguished from random values. Another solution is to have the tag update its ID in some unpredictable fashion (similar to the use of nonces to prevent replay attacks). However, a more general defence against tracking and inventorying is to require that the tag only disclose information to trusted readers, i.e. depends on authentication, which we discuss in the next sub-section. We now consider the question of \emph{where}, i.e. in which layer - application, communication, and physical - attacks against RFID operate.

The classic vulnerability of RFID is that it is wireless and silent; a tag can be read without the knowledge of its owner. Thus, the most common attacks on RFID are unauthorized reading of tags and eavesdropping, i.e. channel attacks. In this context, it is important to remember that though tags may be designed to work with a limited read range (say 10 cm), a powerful rogue reader can cause them to operate at a considerably greater range (50 cm); if it is eavesdropping, i.e. not required to power the tag, the range becomes greater still. When picking up the (stronger) signal from the reader, rather than from a tag, the range of the eavesdropper can be of the order of kilometres. EPC requires that tags choose and send a random bit string to the reader, and further communication is XORed with the bit string, thus protecting against eavesdroppers who can ``hear'' the reader but not the tag. (Better, i.e. more expensive RFID tags can also perform simple symmetric encryption; however, as discussed, it may still be possible to track them without being able to decipher their messages.)

Privacy attacks on RFID could possibly also be executed at the edge hardware, by capturing tags (and possibly readers) and subjecting them to reverse engineering, side-channel or fault analysis. These attacks are designed to breach confidentiality in general, and may be used to forcibly recover an ID from a tag. However, it is unreasonable to expect that the adversary can employ such force without alerting the owner of the object! Hence, such attacks are not usually a privacy concern; they are mostly applied to break authentication (by recovering secret keys etc.)

The backend for RFID consists of standard systems and networks, and standard database and network security principles apply. Transaction histories and identities can be recovered from a compromised backend; secure internet traffic, privacy in databases\cite{L1} and profile management\cite{L36} are essential components of a complete privacy protection system. Ohkubo\cite{Ohkubo03} proposes the concept of \emph{forward security}, i.e. that data transmitted at a specific time cannot be compromised by a future breach in the system.

\subsubsection{Authentication}

RFID, being an identification system, is deeply linked to authentication: identification may be defined as the process of stating one's identity; authentication, of proving it. Besides being originally developed as an authentication system in the military (``identify friend or foe'' systems for fighter aircraft, in WW II), RFID is used in access control systems, payment systems such as credit cards, and in passports etc. 

However, the simple presence of RFID tags is not sufficient to ensure authentication. RFID tags can be writable or reprogrammable; besides, the attacker's replies may not be from a tag at all - Westhues \cite{L59} demonstrate a system to record and play back replies from RFID-based access control systems. Recent examples of compromised RFID authentication include the Dutch passport and Mifare transport card\cite{Mifare}. 

There is also the question (from the previous section on Privacy) of ensuring that the reader is properly authenticated to the tag. Both directions of authentication - tag to reader, and reader to tag - must take into account that an RFID tag is low in computational power (thus incapable of, for example, public-key cryptography).

Attacks on authentication involve an active adversary; it is not possible to break authentication simply by sniffing. Attacks on the communication channel include:
\begin{enumerate}
\item \emph{Impersonation.} In the simplest attack, the adversary simply identifies itself to the server as a tag, and is accepted.  
\item \emph{Replay.} In cases where a tag does not update its responses, old messages between tag and reader can be sniffed by the adversary, and reused for authentication.  \item \emph{Man-in-the-middle.} A specific version, the \emph{relay} attack, is used to defeat schemes that check for physical proximity of a tag; the adversary sets up a back channel between his own ``leech'' reader, which communicates with the tag, and a ``ghost'' that communicates with the reader\cite{Kfir}. More involved MitM attacks modify the messages between tag and reader; it may be possible to use this to corrupt the data on the tag (e.g. marking a bottle of morphine as glucose). 
\item \emph{Race attacks.} Using a stronger or faster transmission than the original reader, the adversary can hijack or (improperly) terminate a session, even after proper authentication between reader and tag.
\item \emph{Noise attacks.} The wireless medium of the signals leaves them open to interference at the physical layer. For example, the Manchester coding for high-speed NFC tags encodes $0$ as a signal at $82\%$ strength, and $1$ as a signal at $100\%$. 
An adversary can boost signal strength; the reader calibrates itself to read a signal at $100\%$ as $0$ and at $125\%$ as $1$ - now the adversary can flip bits.
\end{enumerate}

RFID tags, and to a limited extent readers, are also susceptible to capture and reverse engineering. Common techniques include standard cryptanalysis, side-channel attacks, and \emph{glitching}, in which the system is subjected to abnormal working conditions to cause it to fail. Besides tag ID, which may itself be confidential, such attacks can expose secret keys used for communication, thereby enabling further attacks. Though RFID solutions should ideally be easily computable and resistant to memory analysis, in point of fact, Avoine\cite{Avoine} suggests that it may be a more accurate adversary model to assume that the adversary has access to all on-tag memory, as well as to wireless communication. 

Captured tags are susceptible to \emph{swapping} (attachment to a new object), \emph{cloning}, and \emph{spoofing} (similar to cloning, but without creating a physical new tag) attacks. A more sophisticated adversary can also attempt to attack the integrity of the tag itself - a \emph{reprogramming} attack; in some cases this is possible even when the tag provides no interface for programming, such as by physically altering the state of memory cells\cite{Kuhn98}.

It is not yet common to attack the backend through use of RFID tags and readers, 
using techniques such as buffer overflow and SQL injection, but it is possible \cite{Rieback06}. Such an attack, of course, can take over the entire system and compromise privacy, authentication and availability.

\subsubsection{Availability}

The greatest problem with regard to RFID availability attacks is that it is difficult to diagnose whether there is indeed an active adversary jamming a wireless signal, or there is simple noise, interference with another signal, or passive degradation (due to the presence of metal or water - human bodies can distort or block high-frquency signals). Similarly, an adversary may abuse privacy protection approaches (blocker tags\cite{L32}, RFID guardian\cite{L47}) to cause a Denial of Service attack, or simply shield tags with a layer of metal to form a Faraday cage. Such attacks are usually meant to disrupt the monitoring or alerting system provided by RFID, to enable other attacks such as sabotage, stealing the tagged objects, or possibly simply 
disrupt the functioning of infrastructure.

\subsection{RFID Security Measures}

There are two main approaches to the security of RFID. The first, based on traditional cryptography, presents two challenges - algorithms to perform cryptography on resource constrained devices, and infrastructure to securely manage and deploy such a volume of keys on common objects. The second approach involves physical and policy-based methods. We present a short summary of both types.

\subsubsection{Cryptography}

One primary issue in the use of cryptography for RFID security is the low computational power of RFID tags. More powerful tags can use lightweight variants of AES \cite{L18}, as well as new ciphers using elliptic-curve cryptography \cite{L4}; however, to be usable by simple tags, it is safest to design algorithms that only rely on simple operations such as XOR \cite{L35} or a random number generator \cite{L8}.

The first cryptographic solution for RFID, by Weis et al. \cite{L58}, involved hash locks: when queried, the tag chooses a random value $r$, and returns $r; H(ID, r)$ where $H$ is a secure hash (and $x,y$ means x concatenated with $y$). The reader looks up all possible IDs from a table, and checks which one matches the given hash. (The random number $r$ prevents tracing by an adversary.) 

To reduce the cost of searching through the entire ID space, later solutions allow the reader to share some state with the tag.  Ohkubo\cite{L45} replaces the random component of the hash with a hash-chain, so only $m$ possible replies must be stored.  (The tag has a counter $i$ which starts with the value $m$; each time it talks to a reader, it sends $H^i(ID)$ and reduces $i$ by $1$.) Though forward secure, this scheme is vulnerable to replay attacks. In response, Henrici\cite{L23} keeps track of the count of read attempts since the last successful reader authentication, while Dimitriou\cite{L10} keeps a perfect lock-step synchronization between reader state and tag state, updated with every read.  While these solutions are replay resistant, they are vulnerable to tracking: if a tag spends a long time between reads, its response can still be used to track it.

Juels\cite{L26} brings up the issue that in most practical cases of RFID security, the adversary only has minutes of access to a tag. In such cases, a cheap scheme such as ``a simple list of pseudonyms that cycles to a new ID upon every read request'' is adequate: the attacker, within its limited window, will keep seeing new pseudonyms and be unable either to resolve the ID of the tag, or to track it. However, this scheme does not consider how to share the psedonym-to-ID translation table with genuine readers, and it is possible the adversary may attack the table sharing infrastructure.
Langheinrich and Marti\cite{L39} extend this idea with bit-throttling and shared secrets. Treating the tag ID as a shared secret as per Shamir\cite{L50},  they split it into several pieces (``shares") such that the ID cannot be reconstructed until some threshold number of pieces are known; also, they throttle the tag to reply with a very slow ``trickle'' of bits when queried. The attacker has to spend considerable time to read a tag, while legitimate readers can use caching strategies to quickly find the tag in the set of known tags. 

Another direction in lightweight cryptography is taken by probabilistic protocols, such as Weis' HB+ \cite{L33}. The protocol is one-directional - it only authenticates tags, while hiding ID from eavesdroppers - but its importance lies in the fact that it sets up an NP-hard problem for the adversary, while the tag only needs simple AND and XOR operations. This is possible because the tag ID, say $x$, is a shared secret with the reader. For every challenge $c$ sent by the reader, the tag replies $x.c + y$ where $y$ is noise, with some constant probability $p$, and $0$ otherwise. After a sufficiently large number of rounds $n$, the reader can see that approximately $n(1-p)$ responses match a given $x$, thereby authenticating the tag. The attacker has no way of knowing which of the $n$ responses were noisy, and which were clear.

Molnar and Wagner\cite{L44} propose the idea of a tree-based key-space: Readers keep a tree structure where every node is a key, and every tag holds the keys corresponding to a path in the tree. This general idea is extended by Buttyan \cite{L6}, Dimitriou \cite{L11}, and Lu \cite{L40}. Such approaches are hard to update, as all tags and readers share a key space.

A particularly important challenge to the practical deployment of RFID encryption is ownership transfer, usually involving the updating of keys in an RFID tag to ensure the previous owner of the tagged item (e.g., the supermarket) can no longer access it. Inoue \cite{L25} simply overwrote the original tag ID with a Private ID, storing the original value in a private database.  Osaka\cite{L46} added reader authentication to protect against rewriting attacks, and dynamic pseudonyms to stop tracking attacks. However, this approach still requires a single Trusted Center to translate the pseudonym to the true ID and other information about the tag.

Extending the tree-of-keys approach, Molnar\cite{L43} stores key subtrees on readers with a built-in limit on usage (say, $1000$ authentications); when tag ownership changes, the new owner reads in the tag repeatedly to exhaust the keyspace. A much more ambitious scheme by Berthold et al\cite{L52} advocates hash-locking tags with a consumer-chosen password at checkout using a consumer device that takes over and reprograms tags, a ``data protection card''.

\subsubsection{Distance and Physical Context}

The simplest physical method to force RFID security is to clip tags at checkout; however, this method completely kills the tag, and prevents any further use for them. Karjoth\cite{L34} suggests it is possible to keep the tag, but reduce its range to a few centimeters, by using tear-off antennas. Inoue\cite{L25} suggest using two tags - one with the unique ID to be destroyed on purchase, and a general, low-granularity one to hold information for later use. It may also be simple enough to place the tagged object in a Faraday cage to temporarily disable the tag; aluminium-lined wallets and pouches are built for this very reason.

A more general answer, which works even for embedded tags and objects that cannot be screened, is the ``blocker'' tag\cite{L32}.  This tag responds to all read requests with a jammed signal, so the reader's anti-collision protocol causes it to stop reading. 
With a sufficient number of blocker tags, it is possible to disable readers reading from different orientations and distances. However, blocker tags depend on the adversary reader backing off after it detects the channel is jammed; it may be possible, with differential signal analysis, to separate blocker-tag-only jamming signals from signals where both a blocker-tag and a real tag reply. 

Rieback et al. propose a more powerful solution, a battery-powered, active RFID Guardian\cite{L47}, which ``not only produces a randomly modulated jamming signal, but also allows the user to upload access control lists, indicating which party can perform what operation on which tags''\cite{L}. 

The physical characteristics of the signal can also be used to ensure security. Distance can be measured by signal strength and round-trip time, and orientation, by angle of arrival of the signal. A general design principle is for protocols to have strict timing requirements, and tags to respond immediately, to reduce the attacker's window of opportunity; distance bounding protocols \cite{Hancke08, Clulow06, Preneel05} based on this principle combat relay attacks.

The fact that readers emit radiation (and in many cases, perform a handshake with the tag to authenticate themselves) has also been used as the foundation for physical security. Floerkemeier\cite{L20} propose that RFID standards include ``transparency protocols", such that readers must explicitly broadcast what data they collect, and for whom. Molnar et al\cite{L42} propose that reader devices include trusted computing modules that can attest to their proper functioning. Besides formal auditing, such open disclosure would enable interested users to carry personal devices with ``watchdog tags'' to inspect reader statements. Carrying this idea further, Brainard\cite{L30} propose a TAg Privacy Agent that controls access to personal tags and logs all disclosure of personal data. Kriplean \cite{L37} proposes the concept of physical access control (PAC) where authenticated users can access all RFID data collected in their physical vicinity, based on a map of readers. Clearly, this method can be abused if the attacker places user tags on many agents, and harvests RFID data from multiple physical locations. The authors suggest making it easy to detect tags, with the example of an elevator that mentions the tags of people inside it. However, these physical methods are not secure against eavesdropping (passive) readers, or against rogue readers that do not follow the specification.
\section{Security of Wireless Sensor Networks}

A sensor network consists of small, low-power computing devices, i.e. sensors, deployed within an environment to observe (and report) physical phenomena. Typically, they consist of a considerable number of nodes, which communicate in an ad hoc fashion over wireless channels. Individual nodes, or ``motes'', are highly constrained computational devices - quite similar to RFID tags. However, unlike in the case of RFID, where the focus is on the individual tag (though it may be part of a constellation), the primary focus in WSN is the entire network of sensors.

As for RFID, we consider the security of WSN to have three main objectives: privacy, authentication, and availability.

\subsection{Privacy}

Privacy in a wireless sensor network involves privacy of the data and the query, as well as privacy of context (location etc.)\cite{SKD} Unlike in RFID, it is quite possible that the adversary be \emph{internal} - i.e., be a standard node in the network itself. In this case, it is not sufficient to use encryption or authentication, as the attacker is not masquerading; there must be policies on what data each entity is allowed to know.

The primary constraint on data privacy is that of aggregation. Simply put, a node should not be able to build a more detailed picture of the data than is required for its functioning. For example, if a sensor network is hierarchical, and organized into clusters, the cluster head should be able to report the combined measurement from the sensors, but not their individual readings. The simplest way to do this is for the nodes in a cluster to add noise to their readings; they coordinate with each other over secure channels so that the aggregate (e.g. the sum of their readings) is not affected. This algorithm is called Cluster-based Private Data Aggregation\cite{D14}. In another approach, Slice-Mixed AggRegaTion\cite{D14}, sensors slice up their data and share pieces with their neighbors; the final report contains all the pieces, but jumbled up, so each sensor's reading is kept secret. Or sensors may simply report statistics such as mean, median, range, or the count of data items in various value buckets, instead of reporting the data itself - Generic Privacy-Preservation Solutions\cite{D32}.

The other essential constraint on data privacy is the anonymity of data queries. The usual technique of $k$-anonymity developed from privacy-preserving data mining can be implemented by carrying out $k$ queries for every request (so the observer cannot tell which one was the real query)\cite{D4}; this is, however, expensive. A more involved solution disconnects user from query using tokens; a query is made, not for a particular user, but for the holder of a token, which is purchased from the WSN owner using blind signatures\cite{D37}.

With regard to context privacy, the primary aim is to hide the location of sensors and base stations, and sometimes the time that data was collected (using random delays). The location of the data source can be hidden using flooding, transmissions from fake sources (sensors send fake packets simulating a data source), or random walks, as reported by Zhang\cite{D18}. These basic techniques have been further elaborated in later work. GROW\cite{D28} increases the randomness of the walk: a random walk is carried out from sink to source; when a route is needed, a random walk is carried out from source until it intersects the first walk - hence the path chosen has pieces of two random walks. Proxy-based and Tree-based Filtering Schemes\cite{D29} deal with how to filter out dummy data without breaking source privacy. Finally, similar to Tor, packets can be multiply encrypted (or re-encrypted with each hop), so their appearance changes with each hop\cite{D9} -  though this is usually too computationally expensive for sensor motes.

\subsection{Authentication}

Authentication in Wireless Sensor Networks involves both guarding against spoofing - fake nodes (Sybil attack), fake routes, and fake location - and against unauthorized disclosure of information. Solutions in sensor networks mostly concentrate on cryptography and key management; this is manageable given that sensor networks are small and self-contained, but may be a greater challenge at Internet of Things scale.

In a typical case, a trusted server (usually the base station for the sensor network) acts as a Key Distribution Center and establishes a symmetric key for each sensor node; this key, embedded in the node's memory, is used to authenticate it to the base station. If necessary, the server also generates session keys and sends them to sensor nodes (encrypted with the respective symmetric keys), to allow nodes to communicate in a session. This standard deployment is less secure than ``self-enforcing'' schemes such as Diffie-Hellman key agreement or RSA, as the server may be compromised; however, the limited computational resources of sensor nodes usually do not support public-key cryptography. The recent development of lightweight elliptic-curve cryptography has made it feasible to deploy in some sensor networks, such as those with MICA2 nodes \cite{C9, C10}.

The other main direction in cryptography involves key pre-distribution. Key information is distributed among nodes prior to deployment, and after deployment (which is assumed to be one-time and unplanned), the nodes use this information to authenticate each other. The original idea was proposed by Gligor\cite{Gligor} and extended by various authors\cite{C5, C6}; the current state-of-the-art is the SPINS protocol suite of Perrig\cite{Perrig02}: SNEP for confidential data transfer and $\mu$TESLA for authenticated broadcast.

The second direction of authentication involves the nodes authenticating information they receive, such as time (broadcast from a reference, or shared between nodes) and location (calculated by the distance from reference beacon nodes). Song\cite{D14} demonstrates statistical methods to identify and correct for delayed, replayed, or forged time messages; Du\cite{D18} detects fake location messages by anomaly detection. This attack overlaps with availability; an adversary can attempt to disrupt synchronization or localization with a DoS attack.

Physical attacks on sensor networks usually focus on the edge system; like RFID tags, sensors are usually small, unattended, and easy to capture and tamper with. As sensor nodes are authenticated by their keys, an attacker who controls nodes can clone them and deceive any other nodes that they share pairwise keys with. Besides tamper-proofing, solutions involve effective key management (so that a very large number of nodes have to be compromised to take over the system)\cite{Best}.

\subsection{Availability}

Attacks on availability of Wireless Sensor Networks occur at various layers, but usually target the communication channel.

In the physical layer, the standard attack is to use jamming signals. The standard solutions to jamming, i.e. spread-spectrum (frequency hopping) and use of lower duty cycles for redundancy, apply. Similarly, it is possible to exhaust the communication channel (handled by rate limiting), or cause packet collisions (for which the solution is to employ error-correcting codes). These attacks were studied in detail by Wood\cite{C1}. 

Similar flooding tactics are used at higher layers of the network stack as well. The standard solution, to limit the number of connections by a single sensor (and to make them provide a proof of work, through puzzles)\cite{C3}, has the weakness that it only protects connection-oriented protocols such as TCP, and cannot guard against flooding by ``Hello'' packets. (Using authentication in the handshake performed when setting up a connection will prevent Denial-of-Service through too many connections, but it is possible to overwhelm a node with incomplete requests.) We are not aware of a good solution for connectionless protocols.

Finally, attacks are used to disrupt routing in a network, either by advertising routes and discarding all traffic (blackhole attack), or forwarding only a portion of the traffic (selective forwarding). These attacks were first noted by Wood\cite{C1}, and studied in detail for existing protocols (Directed Diffusion,LEACH) by Wagner\cite{C2}. Such attacks may involve manipulation of the routing information, so their prevention comes under the heading of authentication. In order to survive such attacks, networks must have redundancy and monitoring systems; Du\cite{C19} develop a secure routing protocol for sensor networks.

\section{Concluding Remarks}
Despite the serious interest of the cryptography community and the
networks community, the question of security (and privacy) has few
definitive answers, for RFID as well as for sensor networks. Stumbling
blocks include the wide variety of scenarios where they are deployed;
the fact that tags or motes are usually unattended, and work silently
and wirelessly; and that they are too weak computationally to support
strong cryptographic protocols. However, both RFID and WSN are
critical components of the Internet of Things; with the spread of
automated machine-to-machine configuration and communication, 
there will be a critical need for solutions to the security issues first
encountered in these networks. In this brief survey, we have attempted
to survey the challenges and methods for security in both RFID and
WSN; we hope this will make it easier to spot commonalities and
differences, and perhaps develop a more general security model for the
Internet of Things.

\bibliography{ref}
\bibliographystyle{plain}

\end{document}